\documentclass[10pt]{article}
\usepackage{graphicx}
\newcommand{\no}{\nonumber}

\begin{document}
\begin{titlepage}

\title{\huge Pseudoscalar Form Factors\\ 
       in Tau-Neutrino Nucleon Scattering}
\author{
{}\\{}\\
{\Large Kaoru Hagiwara}\\
{\normalsize \it Theory Group, KEK, Tsukuba 305-0801, JAPAN}\\
{}\\
{\Large Kentarou Mawatari\footnote{{\it E-mail address}: 
                   \texttt{mawatari@radix.h.kobe-u.ac.jp}}}\\
{\normalsize \it Graduate School of Science and Technology,
Kobe University,}\\
{\normalsize \it Nada, Kobe 657-8501, JAPAN}\\ 
{}\\
{\Large Hiroshi Yokoya\footnote{{\it E-mail address}: 
           \texttt{yokoya@theo.phys.sci.hiroshima-u.ac.jp}}}\\
{\normalsize \it Department of Physics, Hiroshima University,}\\
{\normalsize \it Higashi-Hiroshima 739-8526, JAPAN}\\
{\normalsize \it and Radiation Laboratory, RIKEN, Wako 351-0198, JAPAN}
}
\date{}
\maketitle

\begin{abstract}
We investigate the pseudoscalar transition form factors of nucleon 
for quasi-elastic scattering and $\Delta$ resonance production 
in tau-neutrino nucleon scattering via the charged current interactions. 
Although the pseudoscalar form factors play an important role for the 
$\tau$ production in neutrino-nucleon scattering, these are not known well. 
In this article, we examine their effects in quasi-elastic scattering 
and $\Delta$ resonance production and find that the cross section, 
$Q^{2}$ distribution, and spin polarization of the produced 
$\tau^{\pm}$ leptons are quite sensitive to the pseudoscalar form factors.
\end{abstract}

\begin{picture}(5,65)(20,-630)
\put(350,-100){KEK-TH-946}
\put(350,-115){KOBE-FHD-04-01}
\put(350,-130){HUPD-0308}
\put(350,-145){hep-ph/0403076}
\put(350,-160){March 2004}
\end{picture}
\thispagestyle{empty}
\end{titlepage}
\setcounter{page}{1}

Neutrino oscillations in long baseline (LBL) experiments are of great 
interests from both theoretical and experimental point of view.    
It is especially important to confirm $\nu_{\tau}$ appearance 
in LBL experiments in order to demonstrate $\nu_{\mu}\to\nu_{\tau}$ 
oscillation, and $\nu_{\tau}$ should be detected through 
the $\tau$ production by charged current reactions off a nucleon target 
at several LBL neutrino oscillation experiments \cite{barger}, 
such as MINOS \cite{minos}, ICARUS \cite{icarus} and OPERA \cite{opera}. 

As we pointed out in the previous paper \cite{taupol}, the information 
on the spin polarization of $\tau$ produced through the 
neutrino-nucleon scattering is essential to determine the 
$\tau^{\pm}$ production signal since the decay particle distributions 
depend crucially on the $\tau$ polarization.
$\tau$ production followed by its pure leptonic decay 
should also be studied in order to estimate background events 
for the $\nu_{\mu}\to\nu_e$ appearance reactions \cite{aoki}. 
Furthermore, in addition to LBL experiments, the ice/underwater neutrino 
telescopes are expected to detect $\tau$ production in 
neutrino-nucleon scattering, such as at AMANDA \cite{amanda}/IceCube 
\cite{icecube} and Baikal \cite{baikal} experiments.

In LBL experiments, the following three reactions have major 
contribution to the neutrino-nucleon scattering; quasi-elastic 
scattering (QE), resonance production (RES), and deep inelastic 
scattering (DIS). The QE contribution to the $\tau$ production 
dominates the total cross section near threshold, 
$E_{\nu}\sim 3.5$ GeV, and the cross sections of the QE and RES 
processes are significant throughout the energy range of 
$E_{\nu}\simeq 3.5-30$ GeV of the future neutrino oscillation 
experiments \cite{py, taupol}. It is thus important to estimate 
the $\tau$ production cross section and its spin polarization 
for the QE and RES processes.

However, there is an uncertainty in the calculations 
of the cross section and spin polarization of $\tau$ production, 
from the pseudoscalar form factors of those processes.
Because the contribution from the pseudoscalar form factors 
is proportional to the lepton mass, for the $e$ and $\mu$ production case 
these contributions are suppressed and negligible. 
Although there are several experiments which have sensitivity to the 
pseudoscalar form factors, such as in muon capture \cite{mc} 
and in pion electroproduction \cite{pipro}, those results are not
sufficient to constrain these form factors in the range relevant for $\tau$
production \cite{review}. On the other hand, because of the heavy $\tau$ 
mass, $m_{\tau}=1.78$ GeV, the effect of pseudoscalar 
terms to the $\tau$ production can be significant since 
their spin-flip nature is expected to affect the produced $\tau$ 
polarization significantly.

In this letter, we study $\tau$ production in 
the neutrino-nucleon scattering using several 
parameterizations of the pseudoscalar form factors 
in the QE and RES processes, and examine how the production 
cross section and the spin polarization of $\tau$ are affected 
by those form factors.\\

We consider $\tau^-/\tau^+$ production by charged current 
reactions off a nucleon target;
\begin{equation}
\nu_{\tau}(k)/\bar{\nu}_{\tau}(k) + N(p) \to 
\tau^-(k')/ \tau^+(k') + X(p'),
\end{equation}
where the four-momenta are given in brackets and $X$ denotes the final 
hadron. $X$ is a nucleon $N$ for the QE process and $\Delta$ 
or $N+\pi$ for the RES process. 
We define Lorentz invariant variables
\begin{eqnarray}
&&Q^2 = -q^2, \quad  q= k-k',\\
&&W^2 =(p+q)^2,
\end{eqnarray}
where $Q^2$ is the momentum transfer.
Each process is distinguished by the hadronic invariant mass $W$: 
$W=M$ for QE, and $M+m_{\pi}<W<W_{\rm cut}$ for RES. Here 
$W_{\rm cut}$ is an artificial boundary between the RES and 
DIS ($W>W_{\rm cut}$) processes, and we take $W_{\rm cut}=1.6$ GeV
\cite{taupol}.

The $\tau$ production cross section is expressed 
in terms of the leptonic tensor $L^{\mu\nu}$ and the 
hadronic tensor $W_{\mu\nu}$ as 
\begin{equation}
\frac{d\sigma_{\lambda}}{dQ^2\, dW^2}
= \frac{G_F^2\kappa^2}{16\pi M^2 E_{\nu}^2}\,
L^{\mu\nu}_{\lambda}W_{\mu\nu},
\end{equation}
where $G_F$ is Fermi constant, $\kappa =M_W^2/(Q^2+M_W^2)$ is the 
propagator factor with the $W$-boson mass $M_W$, $M$ is the nucleon 
mass, and $E_{\nu}$ is the incoming $\tau$ neutrino energy in the 
laboratory frame. $\lambda$ stands for the produced $\tau$ helicity 
defined in the center-of-mass (CM) frame. 
Explicit form of the leptonic tensor $L^{\mu\nu}_{\lambda}$ in terms of
the $\nu_{\tau}\to\tau^-_{\lambda}$ and 
$\bar{\nu}_{\tau}\to\tau^+_{\lambda}$ transition currents, for the 
$\tau^{\pm}$ helicity $\lambda$ in the CM frame, is found in 
Ref. \cite{taupol}.\\

The hadron tensor for the QE scattering processes
\begin{equation}
\nu_{\tau} + n \to \tau^- +p, \quad
\bar{\nu}_{\tau} + p \to \tau^+ +n,
\end{equation}
is written by using the hadronic weak transition current
$J_{\mu}^{(\pm)}$ as follows \cite{llewellyn}:
\begin{equation}
W^{\rm QE}_{\mu\nu}=\frac{\cos^2\theta_c}{4}\sum_{\rm spins}
J_{\mu}^{(\pm)}{J^{(\pm)}_{\nu}}^{*}\,\delta(W^2-M^2),
\end{equation}
where $\theta_c$ is the Cabibbo angle.
The weak transition currents $J_{\mu}^{(+)}$ and $J_{\mu}^{(-)}$ 
for the $\nu_{\tau}$ and $\bar{\nu}_{\tau}$ scattering, 
respectively, are defined as
\begin{eqnarray}
&&J_{\mu}^{(+)}=\langle p(p')|\hat{J}_{\mu}^{(+)}|n(p)\rangle = 
\bar{u}_p(p')\,\Gamma_{\mu}(p',p)\,u_n(p),\\
&&J_{\mu}^{(-)}=\langle n(p')|\hat{J}_{\mu}^{(-)}|p(p)\rangle = 
\bar{u}_n(p')\,\overline{\Gamma}_{\mu}(p',p)\,u_p(p)=
\langle p(p)|\hat{J}_{\mu}^{(+)}|n(p')\rangle^{*},
\end{eqnarray}
where $\Gamma_{\mu}$ is written in terms of 
the six weak form factors of the nucleon,  
$F^{V}_{1,2,3}$, $F_A$, $F_3^A$ and $F_p$, as
\begin{eqnarray}
\Gamma_{\mu}(p',p)=\gamma_{\mu}\, F^V_1(q^2) \!\!\!&+&\!\!\!
\frac{i\sigma_{\mu\alpha}q^{\alpha}\xi}{2M}\,F^V_2(q^2)
+\frac{q_{\mu}}{M}\,F^V_3(q^2)\no\\
\!\!\!&+&\!\!\!
\left[ \gamma_{\mu}\,F_A(q^2)
+\frac{\left(p+p'\right)_{\mu}}{M}\,F^A_3(q^2)
+\frac{q_{\mu}}{M}\,F_p(q^2) \right]\gamma_5.
\end{eqnarray}
For the $\bar{\nu}_{\tau}$ scattering, the vertex 
$\overline{\Gamma}_{\mu}$ is obtained by 
$\overline{\Gamma}_{\mu}(p',p)=
\gamma_0\Gamma_{\mu}^{\dagger}(p,p')\gamma_{0}$. 

We can drop two form factors, $F^V_3$ and $F^A_3$, because of  
isospin symmetry and time reversal invariance. 
Moreover, the vector form factor $F_1^V$ and $F_2^V$ are related to the 
electromagnetic form factors of nucleons under the conserved 
vector current (CVC) hypothesis:
\begin{equation}
F^V_1(q^2)= {\displaystyle G^V_E(q^2)
-\frac{q^2}{4M^2}G^V_M(q^2) \over 
\displaystyle 1-\frac{q^2}{4M^2}},\quad
\xi F^V_2(q^2)=\frac{\displaystyle G^V_M(q^2)-G^V_E
(q^2)}{\displaystyle 1-{q^2 \over 4M^2}},
\end{equation}
where 
\begin{equation}
G^V_E(q^2)=\frac{1}{\left(1-q^2/M_V^2\right)^2},\quad
G^V_M(q^2)=\frac{1+\xi}{\left(1-q^2/M_V^2\right)^2},
\end{equation}
with a vector mass $M_V=0.84$ GeV and $\xi=\mu_p-\mu_n=3.706$. 
$\mu_p$ and $\mu_n$ are the anomalous magnetic moments of proton and 
neutron, respectively. 
For the axial vector form factor $F_A$, 
\begin{equation}
F_A(q^2)=\frac{F_A(0)}{\left(1-q^2/M_A^2\right)^2}
\end{equation}
with $F_A(0)=-1.267$ and an axial vector mass $M_A=1.0$ GeV.
The above form factors are found to reproduce the $\nu_{\mu}$ and
$\bar{\nu}_{\mu}$ scattering data \cite{review}.

For the pseudoscalar form factor $F_p$, which is the main focus of 
this study, we adopt the following parameterizations with  
different powers of $(1-q^2/M_A^2)$;
\begin{equation}
F_p(q^2)=\frac{2M^2}{m_{\pi}^2-q^2}\,{F_A(0) \over 
\left(1-q^2/M_A^2\right)^n} \quad (n=0,1,2).  \label{fp}
\end{equation}
The normalization of $F_p(0)$ is fixed by the partially conserved 
axial vector current (PCAC) hypothesis. We adopted $n=2$ 
in the previous study \cite{taupol}.

\begin{figure}
\begin{center}
\includegraphics[width=6cm]{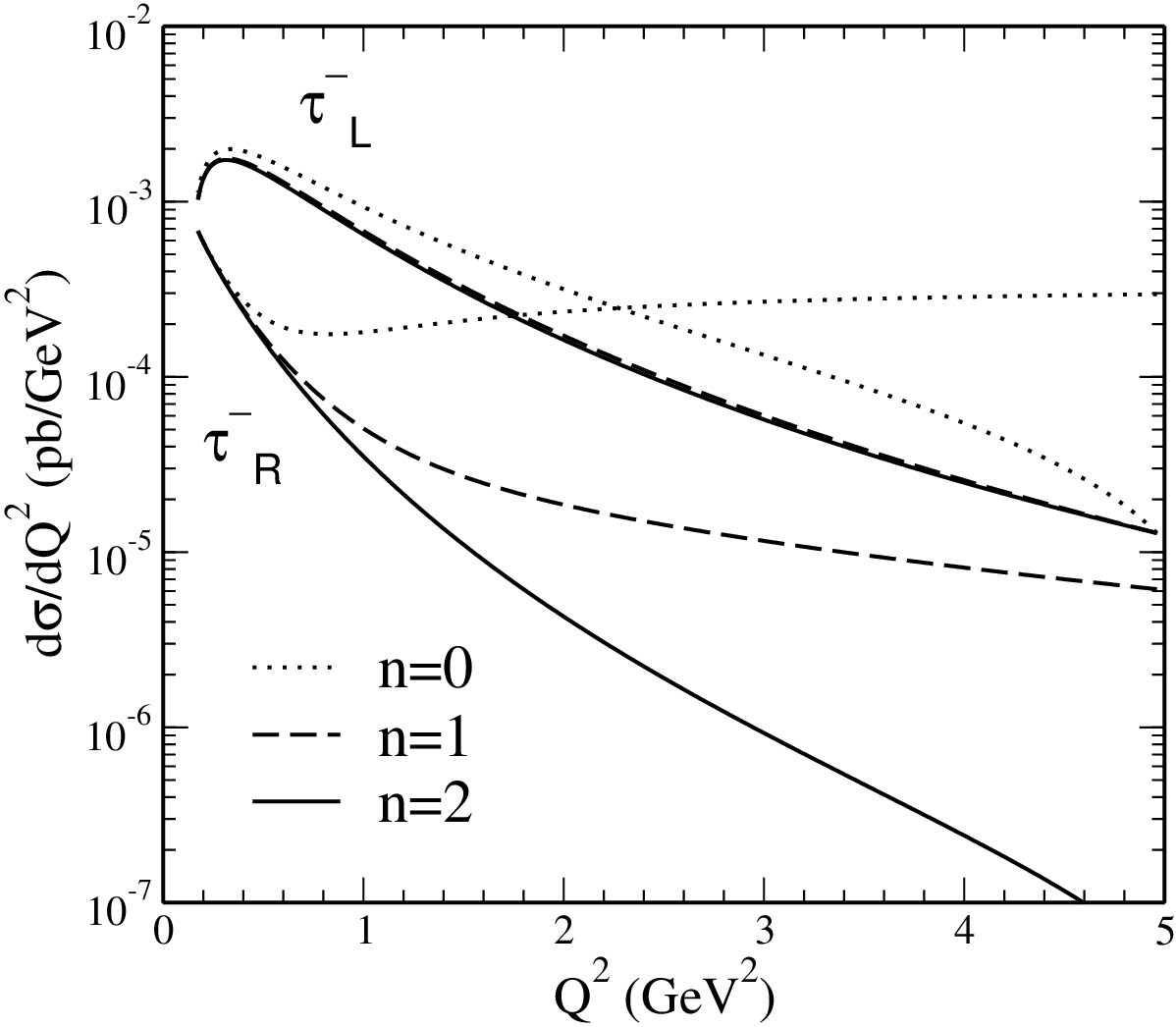} \quad
\includegraphics[width=6cm]{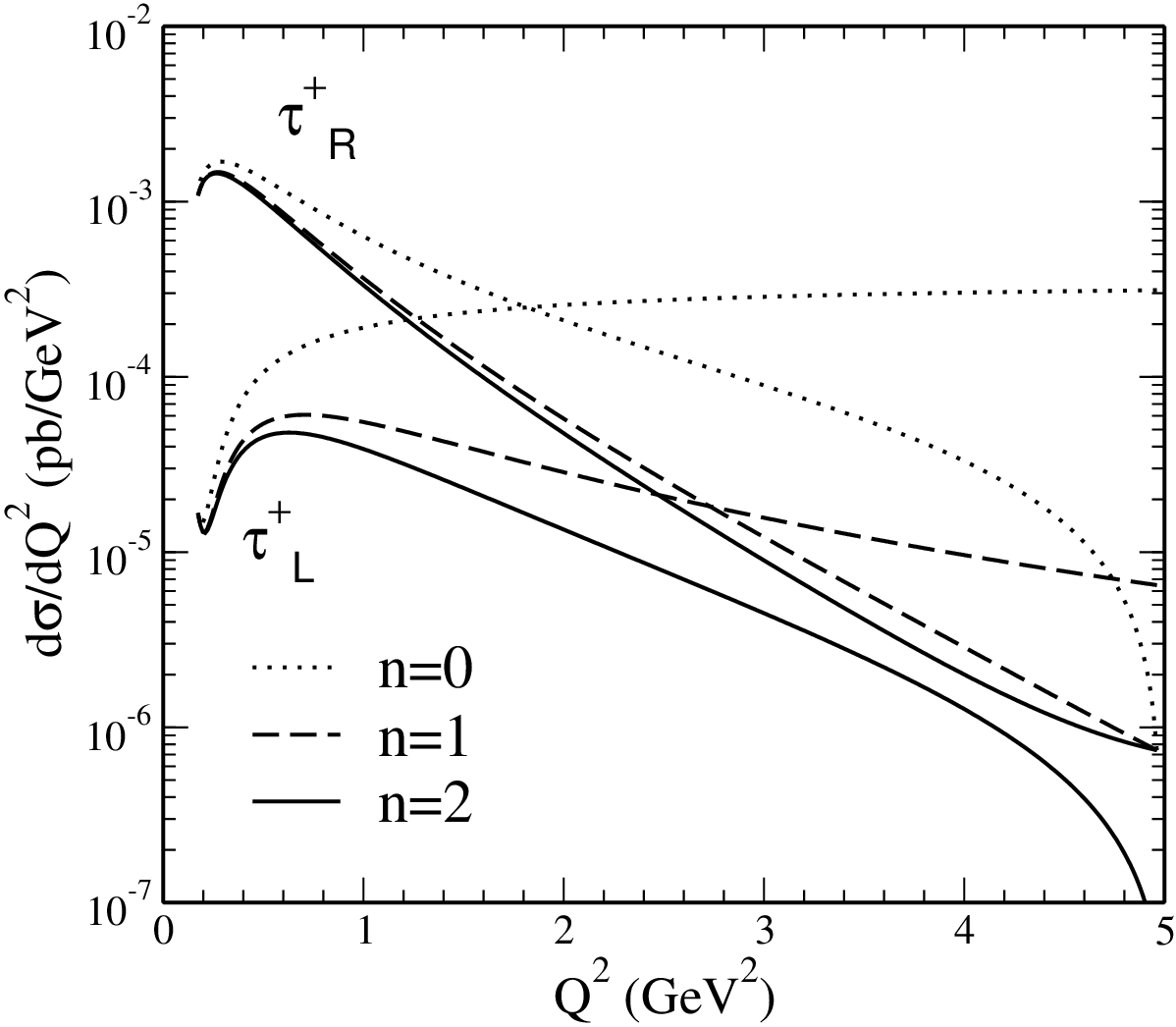}
\end{center}
\caption{
Differential cross sections $d\sigma /dQ^2$ of 
$\tau^-$ (left) and $\tau^+$ (right) productions off the 
isoscalar target in the QE process at neutrino energy $E_{\nu}=5$ GeV. 
The right- and left-handed $\tau$ production are shown separately, where
the helicities are defined in the CM frame.
Solid, dashed, and dotted lines denote $n=2,1,0$, respectively, for the
pseudoscalar form factor Eq. (\ref{fp}).}
\label{qe}
\end{figure} 
In Fig. \ref{qe}, we plot the $Q^2\, (=-q^2)$ dependence of the 
differential cross sections of $\tau$ production off the isoscalar 
target in the QE process at incoming neutrino energy $E_{\nu}=5$ GeV 
in the laboratory frame. 
The left figure is for $\tau^-$ production and the right figure is 
for $\tau^+$ production. 
In the left figure, the upper three lines are for the left-handed 
$\tau^-\ (\tau^-_L)$ production, and the lower three are for 
the right-handed $\tau^-\ (\tau^-_R)$ production. 
Here the helicity is defined in the CM frame.
On the other hand, in the right figure, upper three lines denote 
right-handed $\tau^+\ (\tau^+_R)$ and lower three for left-handed 
$\tau^{+}\ (\tau^+_L)$. 
Solid, dashed, and dotted lines are for $n=2,1,0$, respectively.

We find, while the left-handed $\tau^-$ and the right-handed $\tau^+$
production do not depend much on the pseudoscalar form factor,
the dependences of the right-handed $\tau^-$ and the left-handed $\tau^+$
production on the power of $(1-q^2/M_A^2)$ of the pseudoscalar form
factor are quite significant, especially at large $Q^2$. This feature 
agrees with the spin-flip nature of the pseudoscalar form factor. 
The $n=0$ lines (the pion-pole dominance) give a characteristic 
prediction that the cross sections 
for spin-flipped $\tau$'s ($\tau^-_R$ and $\tau^+_L$) grow at high
$Q^2$. Therefore it should be possible to distinguish between the $n=0$ 
and $n\ge 1$ cases. On the other hand, the difference between the 
$n=1$ and the $n=2$ cases is rather hard to be established  
since the cross section is very small 
in the large $Q^2$ region where the difference becomes large.\\

Next, the hadron tensor for the $\Delta$ production (RES) processes; 
\begin{equation}
\nu_{\tau}+n\ (p) \to \tau^-+\Delta^+\ (\Delta^{++}),\quad
\bar{\nu}_{\tau}+p\ (n) \to
\tau^++\Delta^0\ (\Delta^-),
\end{equation}
is calculated in terms of the nucleon-$\Delta$ weak transition current 
$J_{\mu}$ as follows \cite{llewellyn, schreiner, singh}:
\begin{equation}
W^{\rm RES}_{\mu\nu}=\frac{\cos^2\theta_c}{4}
\sum_{\rm spins}J_{\mu}J^{*}_{\nu}\,
\frac{1}{\pi}\,\frac{W\Gamma(W)}{(W^2-M_{\Delta}^2)^2+W^2\Gamma^2(W)}.
\end{equation}
Here we take the $\Delta$ resonance mass $M_{\Delta}=1.232$ GeV,
and its running width:
\begin{equation}
\Gamma(W) = 
 \Gamma(M_{\Delta})\,\frac{M_{\Delta}}{W}\,\frac{\lambda^{\frac{1}{2}}
(W^2,M^2,m_{\pi}^2)}
{\lambda^{\frac{1}{2}}(M_{\Delta}^2,M^2,m_{\pi}^2)} 
\end{equation}
with $\Gamma(M_{\Delta})=0.12$ GeV and 
$\lambda(a,b,c)=a^2+b^2+c^2-2(ab+bc+ca)$.

The current $J_{\mu}$ for the process 
$\nu_{\tau}+n\to\tau^-+\Delta^+$ is defined by 
\begin{equation}
J_{\mu}=\langle\Delta^+(p')|\hat{J}_{\mu}|n(p) \rangle 
=\bar{\psi}^{\alpha}(p')\,\Gamma_{\mu\alpha}\,u(p),
\end{equation}
where $\psi^{\alpha}$ is the spin-3/2 particle wave function and 
the vertex $\Gamma_{\mu\alpha}$ is expressed in terms of 
the eight weak form factors $C^{V,A}_{i=3,4,5,6}$ as
\begin{eqnarray}
\Gamma_{\mu\alpha}=\left[
\frac{g_{\mu\alpha}\!\not\!q-\gamma_{\mu}q_{\alpha}}{M}
\, C^V_3(q^2)
+\frac{g_{\mu\alpha}\,p'\cdot q-p'_{\mu}q_{\alpha}}{M^2}
\, C^V_4(q^2) 
\right. \no \\ \left.
+\frac{g_{\mu\alpha}\, p\cdot q - p_{\mu}q_{\alpha}}{M^2}
\, C^V_5(q^2)
+\frac{q_{\mu}q_{\alpha}}{M^2}
\, C^V_6(q^2) \right]\gamma_{5}\no\\
+\frac{g_{\mu\alpha}\!\not\!q -\gamma_{\mu}q_{\alpha}}{M}
\, C^A_3(q^2)
+\frac{g_{\mu\alpha}\,p'\cdot q -p'_{\mu}q_{\alpha}}{M^2}
\, C^A_4(q^2) 
\no \\
+g_{\mu\alpha}\, C^A_5(q^2)
+\frac{q_{\mu}q_{\alpha}}{M^2} \, C^A_6(q^2).
\end{eqnarray}
By isospin invariance and the Wigner-Eckart theorem, 
the other nucleon-$\Delta$ weak transition currents 
are given as 
\begin{equation}
\langle\Delta^{++}|\hat{J}_{\mu}|p \rangle = 
\sqrt{3}\langle\Delta^{+}|\hat{J}_{\mu}|n \rangle =
\sqrt{3}\langle\Delta^{0}|\hat{J}_{\mu}|p \rangle =
\langle\Delta^{-}|\hat{J}_{\mu}|n \rangle.
\end{equation}

From the CVC hypothesis, $C^{V}_{6}=0$ and the other vector 
form factors $C^{V}_{i=3,4,5}$ are related to 
the electromagnetic form factors. We adopt the modified dipole  
parameterizations \cite{pys, olsson}:
\begin{equation}
C^V_3(q^2)=\frac{C_3^V(0)}
{\left(1-\displaystyle { q^2\over M_V^2}\right)^2}\,
{ 1\over 1-\displaystyle { q^2\over 4M_V^2}},\quad 
C^V_4(q^2)=-\frac{M}{M_{\Delta}}\,C^V_3(q^2),\quad C^V_5(q^2)=0,
\end{equation}
with $C_3^V(0)=2.05$ and a vector mass $M_V=0.735$ GeV. 

For the axial vector form factors $C^A_{i=3,4,5}$,
several theoretical works were done around 1960--70 
\cite{dennery}--\cite{zuker}.  
Several authors \cite{llewellyn, schreiner} performed 
the comparisons of these predictions in detail with experimental data, 
and showed that the Adler model \cite{adler} modified by Schreiner 
and von Hippel \cite{schreiner} describes the data well at the time. 
However, in face of the recent experimental data \cite{kitagaki}
the $Q^2$ dependence of the weak axial form factors has been re-examined, 
and several authors proposed modified weak axial form factors 
\cite{sato, pys}. We show the several models for $C_5^A$ as examples:
\begin{equation}
C^A_5(q^2) = {C_5^A(0) \over 
\left(1-\displaystyle {q^2\over M_A^2}\right)^2 } \times
\left\{ 
 \begin{array}{ll}
1  & {\rm dipole\ model} \\
{\displaystyle \left(1-\frac{a_A\,q^2}{b_A-q^2}\right)}    
   & {\rm modified\ Adler\ model\ \cite{schreiner}} \vspace*{3pt}\\
{\displaystyle \exp \left[-{a_B\,q^2\over 1-b_B\,q^2}\right]  }  
   & {\rm Bell\ et\ al.\ model\ \cite{bell}} \vspace*{3pt}\\
{\displaystyle (1-a_S\,q^2)\,\exp [b_S\,q^2] }
   & {\rm SL\ model\ \cite{sato}} \vspace*{3pt}\\
\displaystyle {1 \over 1-{q^2 \over 3M_A^2}}
   & {\rm PYS\ model\ \cite{pys} }
  \end{array} \right. \label{c5}
\end{equation}
with $C_5^A(0)=1.2$, an axial vector mass  $M_A=1.0$ GeV.
$a_{A,B,S}$ and $b_{A,B,S}$ are the model 
dependent parameters determined by fitting the experimental data,
$a_{A,B,S}=-1.21,\,-0.61,\,0.154$ and $b_{A,B,S}=2.0,\,0.19,\,0.166$, 
when $q^2$ is measured in units of GeV$^2$. 
For $C_3^A$ and $C_4^A$, $C_3^A=0$ and $C_4^A=-{1\over 4}C_5^A$ give 
good agreements with the data \cite{schreiner}. 
In this report, we adopt the PYS model \cite{pys} in Eq. (\ref{c5}), 
which decrease more rapidly with increasing $Q^2$ than the dipole 
model and the SL model \cite{sato}, and which have more moderate
$Q^2$ dependence as compared to the modified Adler model \cite{schreiner} 
and the Bell et al. model \cite{bell}. 

For the pseudoscalar form factor $C^A_6$, we adopt the same form of the
parameterizations as for the QE case:
\begin{equation}
C^A_6(q^2)=\frac{M^2}{m_{\pi}^2-q^2}\,
{C_5^A(0) \over \left(1-q^2/M_A^2\right)^n} \quad (n=0,1,2) 
\label{c6}
\end{equation}
which agrees with the off-diagonal Goldberger-Treiman 
relation in the limit of $m_{\pi}^2\to 0$ and $q^2\to 0$. 
\begin{figure}[t]
\begin{center}
\includegraphics[width=12cm]{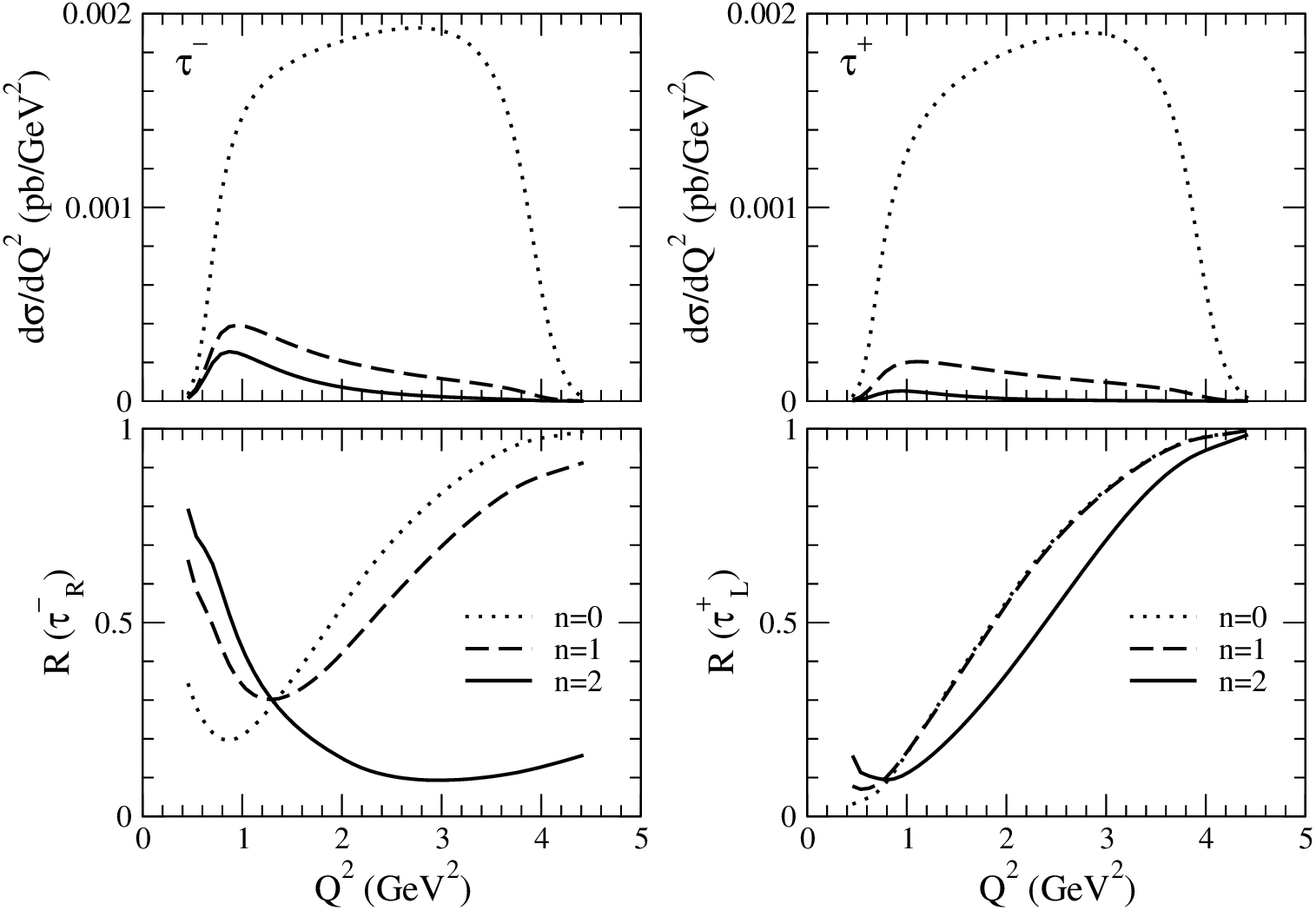}
\end{center}
\caption{
$Q^2$ dependence of the differential cross section (upper) 
and the ratio of the spin-flipped $\tau$ production cross section
(lower) defined in Eq. (\ref{ratio}) 
for $\tau^-$ (left) and $\tau^+$ (right) productions off the 
isoscalar target in the RES process at neutrino energy $E_{\nu}=5$ GeV, 
where the helicities are defined in the CM frame.
Solid, dashed, and dotted lines denote $n=2,1,0$, respectively, 
for the pseudoscalar form factor Eq. (\ref{c6}).}
\label{res}
\end{figure} 

In Fig. \ref{res}, we show the cross section and polarization of 
produced $\tau$ separately. The upper two figures show the $Q^2$ 
dependence of the differential cross section for $\tau^-$ (left figure) 
and $\tau^+$ (right figure) production off the isoscalar target in 
the RES process at neutrino energy $E_{\nu}=5$ GeV.
Solid, dashed, and dotted lines as for $n=2,1,0$, respectively.
Both $\tau^-$ and $\tau^+$ production cross sections  
for $n=0$ are almost 10 times larger than those for $n=1,2$. 
This is because of the absence of extra $Q^2$ suppression to the
pion-pole term in the pseudoscalar form factor. 
Unlike the case for the QE process, the $\Delta$ production cross
sections can distinguish between $n=1$ and $n=2$ cases.

The lower figures show the ratio of the cross section of 
spin-flipped $\tau$ production, 
$\tau^-_R$ (left figure) and $\tau^+_L$ (right figure), defined as
\begin{equation}
R(\tau^-_R)={d\sigma_R \over dQ^2}\bigg/ {d\sigma \over dQ^2},\quad
R(\tau^+_L)={d\sigma_L \over dQ^2}\bigg/ {d\sigma \over dQ^2}, 
\label{ratio}
\end{equation}
where $d\sigma =d\sigma_R + d\sigma_L$.
Here the helicity is defined in the CM frame, as above.
The helicity ratios shown in the lower figures give qualitatively 
different results between $\tau^-$ and $\tau^+$ productions. 
For $\tau^-$, in the large $Q^{2}$ region, the left-handed $\tau^-$ 
dominates for $n=2$, while the right-handed $\tau^-$ dominates for $n=0,1$.
On the other hand, for $\tau^{+}$, the large $Q^{2}$ region is 
dominated by left-handed $\tau^+$ for all $n=0,1,2$ cases. Only
left-handed $\tau^+$ are produced in the backward direction in the CM
frame.

We also examined the parametrization
\begin{equation}
 C_6^A(q^2) = \frac{M^2}{m_{\pi}^2-q^2}\cdot C_5^A(q^2)
\end{equation}
by using the PYS parametrization \cite{pys} of the weak axial vector 
form factor in Eq. (\ref{c5}). 
We find negligible difference from the $n=2$ case, 
for the cross section prediction and the polarization prediction
in the region where the cross section is relatively significant.\\

To summarize, we have studied the pseudoscalar transition 
form factors of nucleon for quasi-elastic scattering and
$\Delta$ resonance production in tau-neutrino nucleon scattering via 
the charged current interactions. 
$Q^2$ dependence of the $\tau^{\pm}$ cross sections was calculated, 
considering the helicities of $\tau$ defined in the CM frame. 
We found that the pseudoscalar form factors significantly enhance  
spin-flip $\tau$ production, right-handed $\tau^-$ and left-handed 
$\tau^+$ production, and that it is possible to determine  
whether those form factors need extra $Q^2$ suppression to the 
pion-pole term or not. $\Delta$ resonance cross sections 
are sensitive to the degree of extra $Q^2$ suppression.

\begin{center}
{\bf \large Acknowledgments}
\end{center}
\hspace*{\parindent}
We are grateful to M. Sakuda for useful comments, discussions, 
and encouragements.
K.M. and H.Y. thank the KEK theory group for its hospitality, 
where a part of this work was performed. 
K.M. would like to thank T. Morii for discussions.
H.Y. would like to thank M. Hirata, J. Kodaira for discussions, 
and RIKEN BNL Research Center for its hospitality, where a part
of this work was performed. 


\end{document}